\documentclass[useAMS,usenatbib]{lib/mnras}
\usepackage{natbib}
\usepackage{epsfig}
\usepackage{scrtime}
\usepackage{graphicx}	
\usepackage{amsmath}	
\usepackage{amssymb}	
\usepackage{float}
\usepackage[caption = false]{subfig}
\usepackage{multirow}
\usepackage{rotating}
\usepackage{ulem}
\usepackage{booktabs}
\usepackage{caption} 
\usepackage{threeparttable} 
\usepackage{threeparttablex}
\usepackage{fancyhdr} 
\usepackage{longtable}
\usepackage{xcolor}
\usepackage{ulem}
\usepackage{multibib}

\newcommand{\logd}{\log}
\newcommand{\hii}{H\,\textsc{ii}}

\newcommand{\lsigG}{$L - \sigma$}

\newcommand{\mincir}{\raise-3.truept\hbox{\rlap{\hbox{$\sim$}}\raise4.truept\hbox{$<$}\ }}

\title[Mapping the Hubble Flow with HII Galaxies]{ Mapping the Hubble Flow from z$ \sim $0 to z$ \sim$7.5 with HII Galaxies}
\author[Ch\'avez et al.] {R.\, Ch\'avez$^{1, 2}$\thanks{E-mail: r.chavez@irya.unam.mx}, 
R.\, Terlevich$^{3, 4, 5}$, 
E.\, Terlevich$^{3, 4, 5}$, 
A.\ L.\, Gonz\'alez-Mor\'an$^{3, 6}$, 
D.\, Fern\'andez-Arenas$^{7, 1, 3}$, \newauthor
F.\, Bresolin$^{8}$,
M.\, Plionis$^{9, 10, 11}$,
S.\, Basilakos$^{12, 13, 14}$, 
R.\, Amorín$^{15}$ and 
M.\, Llerena$^{16}$ 
\\ \\
$^{1}$Universidad Nacional Autónoma de México, Instituto de Radioastronomía y Astrofísica, 58090, Morelia, Michoacán, México \\
$^{2}$Secretaría de Ciencia, Humanidades, Tecnología e Innovación, Av. Insurgentes Sur 1582, 03940, Ciudad de México, México \\
$^{3}$Instituto Nacional de Astrof\'\i sica, \'Optica y Electr\'onica,Tonantzintla, Puebla, M\'exico \\
$^{4}$Institute of Astronomy, University of Cambridge, Cambridge, CB3 0HA, UK \\
$^{5}$Facultad de Astronom\'\i a y Geof\'\i sica, Universidad de La Plata, La Plata, Argentina \\
$^{6}$Universidad Pedag\'ogica del Estado de Sinaloa, Culiac\'an, Sinaloa, 80027, M\'exico \\
$^{7}$Canada–France–Hawaii Telescope, Kamuela, 96743 HI , USA \\
$^{8}$Institute for Astronomy, University of Hawaii, 2680 Woodlawn Drive, 96822 Honolulu,HI USA \\
$^{9}$National Observatory of Athens, Lofos Nymfon, 11852 Athens, Greece \\
$^{10}$Physics Dept., Aristotle Univ. of Thessaloniki, Thessaloniki 54124, Greece \\
$^{11}$CERIDES, Center of Excellence in Risk \& Decision Sciences, European University of Cyprus, Cyprus \\
$^{12}$National Observatory of Athens, P.Pendeli, Athens, Greece \\
$^{13}$Academy of Athens, Research Center for Astronomy and Applied Mathematics, Soranou Efesiou 4, 11527, Athens, Greece \\
$^{14}$School of Sciences, European University Cyprus, Diogenes Street, Engomi, 1516 Nicosia, Cyprus \\
$^{15}$Instituto de Astrof\'{i}sica de Andaluc\'{i}a (CSIC), Apartado 3004, 18080 Granada, Spain \\
$^{16}$INAF - Osservatorio Astronomico di Roma, Via di Frascati 33, 00078, Monte Porzio Catone, Italy
}

\begin{document}

\date{MN-24-2288-MJ.R1 --- Compiled at \thistime\ hrs  on \today\ }

\pagerange{\pageref{firstpage}--\pageref{lastpage}} \pubyear{2024}

\maketitle

\label{firstpage}

\begin{abstract}
Over twenty years ago, Type Ia Supernovae (SNIa) observations revealed an accelerating Universe expansion, suggesting a significant dark energy presence, often modelled as a cosmological constant, \( \Lambda \). Despite its pivotal role in cosmology, the standard $\Lambda$CDM model remains largely underexplored in the redshift range between distant SNIa and the Cosmic Microwave Background (CMB). This study harnesses the James Webb Space Telescope's advanced capabilities to extend the Hubble flow mapping across an unprecedented redshift range, from \( z \approx 0 \) to \( z \approx 7.5 \). Using a dataset of 231 HII galaxies and extragalactic HII regions, we employ the \(\text{L}-\sigma\) relation that correlates the luminosity of Balmer lines with their velocity dispersion, to define a competitive technique for measuring cosmic distances. This approach allows the mapping of the Universe
expansion history over more than 12 billion years, covering 95\% of its age. Our analysis, using Bayesian inference, constrains the parameter space $\lbrace h, \Omega_m, w_0\rbrace = \lbrace 0.731\pm0.039, 0.302^{+0.12}_{-0.069}, -1.01^{+0.52}_{-0.29}\rbrace $ (statistical) for a flat Universe. Our results provide new insights into cosmic evolution and imply a lack of change in the photo-kinematical properties of the young massive ionizing clusters in  HII galaxies across most of the history of the Universe.
\end{abstract}


\section{Introduction}\label{sec1}
The first conclusive evidence supporting the accelerated expansion of the Universe was presented over two decades ago through the analysis of Type Ia Supernovae (SNIa) data \citep{Riess1998, Perlmutter1999}. Subsequent studies involving cosmic microwave background (CMB) anisotropies \citep[e.g.][]{Jaffe2001, Pryke2002, Spergel2007, PlanckCollaboration2014, PlanckCollaboration2016a} and Baryon Acoustic Oscillations (BAOs) \citep[e.g.][]{Eisenstein2005, Blake2011}, along with independent measurements of the Hubble parameter \citep[e.g.][]{Chavez2012, Freedman2012, Riess2016, Riess2018, Fernandez2018}, have robustly confirmed the existence of a dark energy (DE) component in the Universe.

Observational campaigns targeting high redshift (\( z \)) cosmological tracers are underway to refine our understanding of the DE Equation of State (EoS). A central objective is to ascertain if the \( w \) parameter, which determines the relationship between pressure \( p \) and the mass-energy density \( \rho c^2 \) within the DE EoS, undergoes evolution over cosmic time \citep{PeeblesRatra1988, Wetterich1988}. By rigorously constraining cosmological parameters and cross-validating findings via diverse, independent methodologies, we aim to produce a more accurate and resilient cosmological framework.

\hii\ galaxies (HIIGs) represent intense and compact star formation episodes predominantly found in dwarf irregular galaxies, where they significantly contribute to the overall luminosity. HIIGs are spectroscopically selected as the star forming systems with the largest equivalent width of their Balmer emission lines ($EW(\mathrm{H}\beta) > 50$ \AA). This selection 
criterion guarantees that HIIGs are the youngest systems ($\lesssim$ 5 Myr) that can be studied in detail. Similarly, giant extragalactic \hii\ regions (GEHRs) are associated with vigorous star formation, but they are typically situated in the outer disks of late-type galaxies. Notably, the rest-frame optical spectra of both GEHRs and HIIGs are virtually indistinguishable from each other, marked by pronounced emission lines. These lines arise from gas ionisation events triggered by a massive Young Stellar Cluster (YSC) or an ensemble of such clusters, often referred to as a Super Star Cluster (SSC) \citep{Searle1972, Bergeron1977, Terlevich1981, Kunth2000, Chavez2014}.

The pronounced emission lines in the rest-frame optical spectra of GEHRs and HIIGs position them as potent probes for investigating young star formation at high-\(z\). With instruments such as NIRSpec \citep{Dorner2016} onboard the JWST \citep{Gardner2006}, it becomes feasible to delve into these regions up to \(z \sim 6.5\) via the H\(\alpha\) emission line or even up to \(z \sim 9\) using the H\(\beta\) and \mbox{[\(\text{OIII}\)]\(\lambda\lambda4959,5007\) \AA} emission lines. Consequently, this allows for the observation of luminous HIIGs, tracing back to and including the epoch of reionisation.

Multiple studies have proven that both HIIGs and GEHRs exhibit a correlation between their Balmer lines luminosity, e.g. \(L(H\beta)\), and the ionised gas velocity dispersion, \(\sigma\), traced by these emission lines. This correlation, known as the \(\text{L}-\sigma\) relation \citep{Terlevich1981, Melnick_Terlevich_Moles1988, Bordalo_Telles2011, Chavez2014}, serves as a powerful cosmological distance indicator \citep{Plionis2011, Chavez2016, GonzalezMoran2019, 2021MNRAS.505.1441G}, where GEHRs and nearby HIIGs are used as the “anchor” sample because their distances can be independently estimated from Cepheid variables or Tip of the Red Giant Branch (TRGB) measurements \citep{Chavez2012, Fernandez2018}. Consequently, the \(\text{L}-\sigma\) relation offers a unique avenue for employing this distance estimator to study the Hubble flow across a vast \(z\) range.

In this work we present the use of the \(\text{L}-\sigma\) relation of GEHRs and HIIGs as a cosmological tracer up to $z \sim 7.5$ and the resulting constraints to cosmological parameters. Our analysis demonstrates the efficacy of GEHRs and HIIGs in tracing the evolution of the Universe, offering insights into the dynamics of the cosmic expansion and the distribution of matter across a significant fraction of cosmic history. The results presented herein covering an unprecedented redshift range, from 0.0 to $\sim$ 7.5, contribute to the ongoing efforts to constrain cosmological models with greater precision, utilising the unique properties of these astrophysical objects.

\section{Data sets}
The dataset employed in this study is derived from a comprehensive compilation  sourced from multiple previously published works:
\begin{itemize}
    \item The anchor sample of 36 nearby objects with independently measured distance moduli was presented and studied in \citet{Fernandez2018}  and references therein.
    \item At low $z$ ($0.01 \leq z \leq 0.15$), we use 107  HIIGs  extensively analysed in \citet{Chavez2014}.
    \item At intermediate $z$ ($0.6 \leq z \leq 4$) we included 24 HIIGs from \citet{Erb2006}, \citet{Masters2014} and \citet{Maseda2014}, our 6 HIIGs observed with VLT-XShooter published in \citet{Terlevich2015}, 15 HIIGs observed with Keck-MOSFIRE and presented in \citet{GonzalezMoran2019} and 29 HIIGs observed with VLT-KMOS and presented in \citet{2021MNRAS.505.1441G}. Here we also include 9 new HIIGs first presented in \citet{2023A&A...676A..53L}.
    \item At high $z$ ($4 \leq z \leq 7.5$) we include 5 new HIIGs observed with JWST-NIRSpec as part of the JWST Advanced Deep Extragalactic Survey (JADES) \citep{2024A&A...690A.288B} and first presented in \citet{2024A&A...684A..87D}. 
\end{itemize}

The newly collected data for HIIGs, which extends our previously published data is given in Table \ref{tab:data}.  This dataset integrates the 5 HIIGs observed through JWST-NIRSpec, as described in \citet{2024A&A...684A..87D}, along with 9 HIIGs from \citet{2023A&A...676A..53L}. Recognising the well-documented disparity in  velocity dispersion measurements ($\sigma$) measured either with the [OIII] or the Balmer lines, a correction of $2.1\ \mathrm{km\ s^{-1}}$ \citep{Chavez2016} derived from data with both measurements, was applied to three JWST-NIRSpec HIIGs which do not have $\sigma$ measured from Balmer emission lines. Additionally, we performed extinction corrections on all the H$\beta$ flux measurements, following the extinction law in \citet{Gordon2003}. The Balmer decrement, derived from H$\alpha$ and H$\beta$ fluxes, was primarily used in the JWST-NIRSpec data  and most of the VUDS/VANDELS dataset \citep{2023A&A...676A..53L} for extinction correction. Where  H$\alpha$ fluxes were not available, we applied the mean extinction value from the  VUDS/VANDELS 
data. 

\begin{table*}
        \caption{Data set measurements used in the analysis.}
\label{tab:data}
\begin{tabular}{@{}lcccc@{}}
\toprule
Object & $z$ & $\logd \sigma$ & $\logd f (\mathrm{H}\beta)$ & $EW (\mathrm{H}\beta)$ \\
&&($\mathrm{km\ s^{-1}}$)&($\mathrm{erg\ s^{-1}\ cm^{-2}}$)&($\mathrm{\AA}$)\\
\midrule
\multicolumn{5}{c}{\textbf{JWST data}} \\
JADES-NS-00016745 & $5.56616 \pm 0.00011^{a}$ & $1.731 \pm 0.016^{b}$& $-17.64 \pm 0.21^{d}$ & $135.35 \pm 10.47^{f}$ \\
JADES-NS-10016374 & $5.50411 \pm 0.00007^{a}$ & $1.785 \pm 0.014^{b}$& $-18.14 \pm 0.20^{d}$ & $212.74 \pm 12.53^{f}$ \\
JADES-NS-00019606 & $5.88979 \pm 0.00008^{a}$ & $1.622 \pm 0.019^{c}$& $-18.11 \pm 0.27^{d}$ & $242.15 \pm 48.25^{f}$ \\
JADES-NS-00022251 & $5.79912 \pm 0.00007^{a}$ & $1.621 \pm 0.011^{c}$& $-17.86 \pm 0.11^{d}$ & $208.52 \pm 9.64^{f}$\\
JADES-NS-00047100 & $7.43173 \pm 0.00015^{a}$ & $1.868 \pm 0.024^{c}$& $-17.62 \pm 0.39^{e}$ & \\
\midrule
\multicolumn{5}{c}{\textbf{VUDS data}} \\
5101421970  & $2.4710 \pm 0.00025^{g}$& $1.706 \pm 0.022^{g}$ & $-16.35 \pm 0.14^{h}$ & $~49.72 \pm 3.01^{h}$ \\
510996058   & $2.4935\pm 0.00025^{g}$& $1.740 \pm 0.094^{g}$ & $-16.62 \pm 0.42^{h}$ & $~50.55 \pm 13.64^{h}$\\
511001501   & $2.2247\pm 0.00022^{g}$& $1.779 \pm 0.028^{g}$ & $-16.05 \pm 0.08^{h}$ & $197.56\pm 10.72^{h}$\\
5101444192  & $3.4205\pm 0.00034^{g}$& $1.845 \pm 0.017^{g}$ & $-16.31 \pm 0.16^{h}$ & $144.67\pm 25.55^{h}$\\
\midrule
\multicolumn{5}{c}{\textbf{VANDELS data}} \\
UDS022487& $3.0679\pm 0.00031^{g}$& $1.710 \pm 0.031^{g}$ &
$-16.55 \pm 0.15^{h}$ & $85.33 \pm 8.95^{h}$ \\
CDFS020954&$3.4993\pm 0.00035^{g}$& $1.761 \pm 0.085^{g}$ &
$-16.42 \pm 0.14^{h}$&$152.87\pm 8.2^{h}$  \\
CDFS022799&$2.5457\pm 0.00025^{g}$& $1.787 \pm 0.017^{g}$ & $-16.31 \pm 0.07^{h}$ & $~~80.3  \pm 3.83^{h}$ \\
UDS020394& $3.3076\pm 0.00033^{g}$& $1.842 \pm 0.014^{g}$ & $-16.72 \pm 0.15^{h}$ & $125.37\pm 19.01^{h}$\\
CDFS018182&$2.3174\pm 0.00023^{g}$& $1.850 \pm 0.016^{g}$ & $-16.37 \pm 0.08^{h}$ & $50.43 \pm 2.57^{h}$ \\
\hline
\multicolumn{5}{l}{}\\
\multicolumn{5}{l}{$^{a}$Taken from \cite{2024A&A...684A..87D}. $^b$Measured from the H$\alpha$ line in \citet{2024A&A...684A..87D} and corrected by}\\
\multicolumn{5}{l}{thermal broadening. $^c$Measured from the [OIII] line in \citet{2024A&A...684A..87D} and corrected by thermal} \\
\multicolumn{5}{l}{  broadening then corrected to the Balmer lines value (see the text). $^d$Taken directly from the JADES DR2} \\
\multicolumn{5}{l}{  catalogue \citep{2024A&A...690A.288B} and corrected by extinction (see the text). $^e$Obtained from \citet{2025NatAs...9..141B}} \\
\multicolumn{5}{l}{ and corrected for extinction. $^f$Measured directly from JADES spectra. $^g$M. Llerena, personal communication.} \\
\multicolumn{5}{l}{  $^h$Taken from \citet{2023A&A...676A..53L}.}
\end{tabular}
\end{table*}

\section{Constraints on Cosmological parameters}
To rigorously define the cosmological parameters within the scope of this study, we employ a refined methodology, building upon the foundational approach delineated in our preceding works \citep{Chavez2016, GonzalezMoran2019}. A succinct overview of this methodology is presented below to facilitate a comprehensive understanding.

The likelihood function used for the analysis of GEHRs and HIIGs is expressed as:
\begin{equation}
    \mathcal{L}_{H} \propto \exp{\left(- \frac{1}{2} \chi^2_{H}\right)}\;,
    \label{eq:lkh}
\end{equation}
where the chi-squared ($\chi^2_{H}$) term is defined by:
\begin{equation}
    \chi^2_{H} =  \sum_{n}\frac{\left(\mu_o(\log f, \log \sigma | \alpha, \beta) - \mu_{\theta}(z | \theta)\right)^2}{\epsilon^2}\;.
    \label{eq:chi}
\end{equation}
Here $\mu_o$ represents the `observed' distance modulus, derived from the observables via the \(\text{L}-\sigma\) relation, 
\begin{equation}
    \mu_o = 2.5(\beta \log \sigma + \alpha - \log f - 40.08),
\end{equation}
where $\alpha$ and $\beta$ are respectively  the \lsigG\ relation's intercept and slope, $\log \sigma$ is the logarithm of the velocity dispersion,
corrected for broadening (thermal and instrumental), and $\log f$ is the
logarithm of the extinction corrected flux. In the other hand $\mu_{\theta}$ denotes, for HIIGs, the distance modulus derived from a cosmological model with parameters $\theta$ and the measured redshift $z$, while for our anchor sample it represents the distance modulus measured via a primary distance indicator. 

In Equation~\ref{eq:chi}, the theoretical distance modulus, $\mu_{\theta}$, is a function of a set of cosmological parameters. In the broadest scenario considered in this study, these parameters are denoted as $\theta = \{h, \Omega_m, w_0, w_a\}$, in addition to the redshift, $z$. The parameters $w_0$ and $w_a$ are pivotal in defining the DE EoS. Its general form  is given by:
\begin{equation}
    p_{w} = w(z) \rho_{w} c^2,
    \label{eq:DE_EoS}
\end{equation}
where $p_{w}$ represents the pressure, and $\rho_{w}$ denotes the density of  DE. The function $w(z)$ characterises the evolving DE EoS parameter. Various DE models have been proposed and explored, many of which employ a Taylor expansion around the present epoch. A notable example is the Chevallier-Polarski-Linder (CPL) parametrization \citep{Chevallier2001, Linder2003, Peebles2003, Dicus2004, Wang2006}, which is expressed as:
\begin{equation}
    w(z) = w_0 + w_a \frac{z}{1 + z}.
    \label{eq:CPL_model}
\end{equation}
The cosmological constant, denoted as $\Lambda$, is just a special case of DE, given for $(w_0,w_a)=(-1,0)$  while the so called wCDM models are such that $w_a=0$ but $w_0$ can take values $\neq -1$.

In the likelihood function, the weights are quantified by $\epsilon^2$, which encapsulates various sources of uncertainties. This is formally represented as:
\begin{equation}
    \epsilon^2 = \epsilon^2_{\mu_{o}, \text{stat}} + \epsilon^2_{\mu_{\theta}, \text{stat}} + \epsilon^2_{\text{sys}},
    \label{eq:epsilon}
\end{equation}
where $\epsilon_{\mu_{o}, \text{stat}}$ denotes the statistical uncertainties of the observed distance modulus, defined by:
\begin{equation}
    \epsilon^2_{\mu_{o}, \text{stat}} = 6.25 \left( \epsilon_{\log f}^2 + \beta^2 \epsilon_{\log \sigma}^2 + \epsilon_{\beta}^2 \log \sigma^2 + \epsilon_{\alpha}^2 \right).
    \label{eq:epsilon_mu_o_stat}
\end{equation}
Here, $\epsilon_{\log f}$, $\epsilon_{\log \sigma}$, $\epsilon_{\alpha}$, and $\epsilon_{\beta}$ represent the uncertainties associated with the logarithm of the flux, the logarithm of the velocity dispersion, and the intercept and slope of the \(\text{L}-\sigma\) relation, respectively. Furthermore, $\epsilon_{\mu_{\theta}, \text{stat}}$ in Equation~\ref{eq:epsilon} refers to the statistical uncertainty associated with the theoretical distance modulus. This uncertainty originates from the redshift uncertainty in the case of HIIGs, and from the primary distance indicator measurement uncertainty for the anchor sample. Lastly, $\epsilon_{\text{sys}}$ encompasses the systematic uncertainties.

In the pursuit of a more versatile analysis framework, we have also established an \( h \)-free likelihood function, as suggested by \cite{Nesseris2005}. This involves a rescaling of the luminosity distance (\( d_L \)) through the introduction of a dimensionless luminosity distance, \( D_L(z,\theta) \), defined as:
\begin{equation}
D_L(z,\theta)= (1+z) \int_{0}^{z}{\frac{dz^{'}}{E(z^{'}, \theta)}}
\end{equation}
In this formulation, \( d_L \) is expressed as \( d_L = c D_L / H_0 \). This rescaling technique is employed to ascertain cosmological parameters independently of the Hubble constant, a methodology comprehensively detailed in \cite{GonzalezMoran2019}. Here $E(z, \theta)$ for a flat Universe is given by:
\begin{equation}\label{eq:Ez}
        E^2(z, \theta) = \Omega_{r}(1+z)^4 + \Omega_{m}(1+z)^3 + \Omega_{w} (1+z)^{3y}\exp\left( \frac{-3 w_a z}{1+z}\right)
\end{equation}
with $y=(1 + w_0 + w_a)$ and $\Omega_{r}$ the radiation density parameter, such that we can define $\Omega_{w} = 1 - \Omega_{m} - \Omega_{r}$.

In our analysis, we employ the MultiNest Bayesian inference algorithm \citep{Feroz2008, Feroz2009, 2019OJAp....2E..10F} to optimise the likelihood function, thereby deriving constraints on various combinations of nuisance and cosmological parameters. Uniform uninformative priors are consistently used in all cases \citep[cf.][]{2021MNRAS.505.1441G}.

In order to facilitate a comprehensive comparison of our derived constraints with existing studies, we have adopted the figure of merit (\(FoM\)) as defined by \citet{Wang2008}. This \(FoM\) is quantitatively expressed as:
\begin{equation}
	FoM = \frac{1}{\sqrt{\text{det}\ \text{Cov}(\theta_0, \theta_1, \theta_2, \ldots)}}
\end{equation}
where, \(\text{Cov}(\theta_0, \theta_1, \theta_2, \ldots)\) represents the covariance matrix corresponding to the parameter set \(\{\theta_i\}\). This metric provides a robust quantitative basis for evaluating and comparing the precision of different cosmological parameter estimations. 


\begin{table*}
        \caption{Marginalised best-fit parameter values and associated $1\sigma$ uncertainties for the HIIGs and anchor samples. Values enclosed in parentheses indicate parameters that were held constant during the analysis.}
\label{tab:const}
\begin{tabular}{@{}lccccccc@{}}
\toprule
	Data Set       & $\alpha$ & $\beta$ & $h$ & $\Omega_m$ & $w_0$ & $w_a$  & N\\
\midrule
	HIIG & --- & ($5.022\pm 0.058$) & --- & $0.282^{+0.037}_{-0.045}$ & (-1.0) & (0.0) & 195\\
	HIIG & --- & ($5.022\pm 0.058$) & --- & $0.278^{+0.092}_{-0.051}$ & $-1.21^{+0.45}_{-0.40} $ & (0.0) & 195\\
\midrule
	HIIG & ($33.268\pm 0.083$) & ($5.022\pm 0.058$) & $0.715\pm 0.018$ & $0.267^{+0.038}_{-0.048}$ & (-1.0) & (0.0) & 195\\
	HIIG & ($33.268\pm 0.083$) & ($5.022\pm 0.058$) & $0.718\pm 0.020$ & $0.278^{+0.091}_{-0.050}$ & $-1.22^{+0.46}_{-0.40}$ & (0.0) & 195\\
\midrule
    Anchor+HIIG & $33.276\pm 0.110$ & $4.997\pm 0.089$ & $0.730\pm 0.038$ & (0.3) & (-1.0) & (0.0) & 231\\
	Anchor+HIIG & $33.276\pm 0.138$ & $4.997\pm 0.113$ & $0.730\pm 0.040$ & $0.335^{+0.044}_{-0.055}$ & (-1.0) & (0.0) & 231 \\ 
	Anchor+HIIG & $33.285\pm 0.138$ & $4.989\pm 0.113$ & $0.731\pm 0.039$ & $0.302^{+0.12}_{-0.069}$ & $-1.01^{+0.52}_{-0.29}$ & (0.0) & 231\\ 
    Anchor+HIIG & $33.290\pm 0.137$ & $4.986\pm 0.112$ & $0.730\pm 0.039$ & $0.321^{+0.10}_{-0.063}$ & $-0.91^{+0.55}_{-0.33}$ & $-0.71^{+0.65}_{-1.2}$ & 231\\ 
\hline
\end{tabular}
\end{table*}

\begin{figure*}
\begin{center}
\includegraphics[width=1.6\columnwidth]{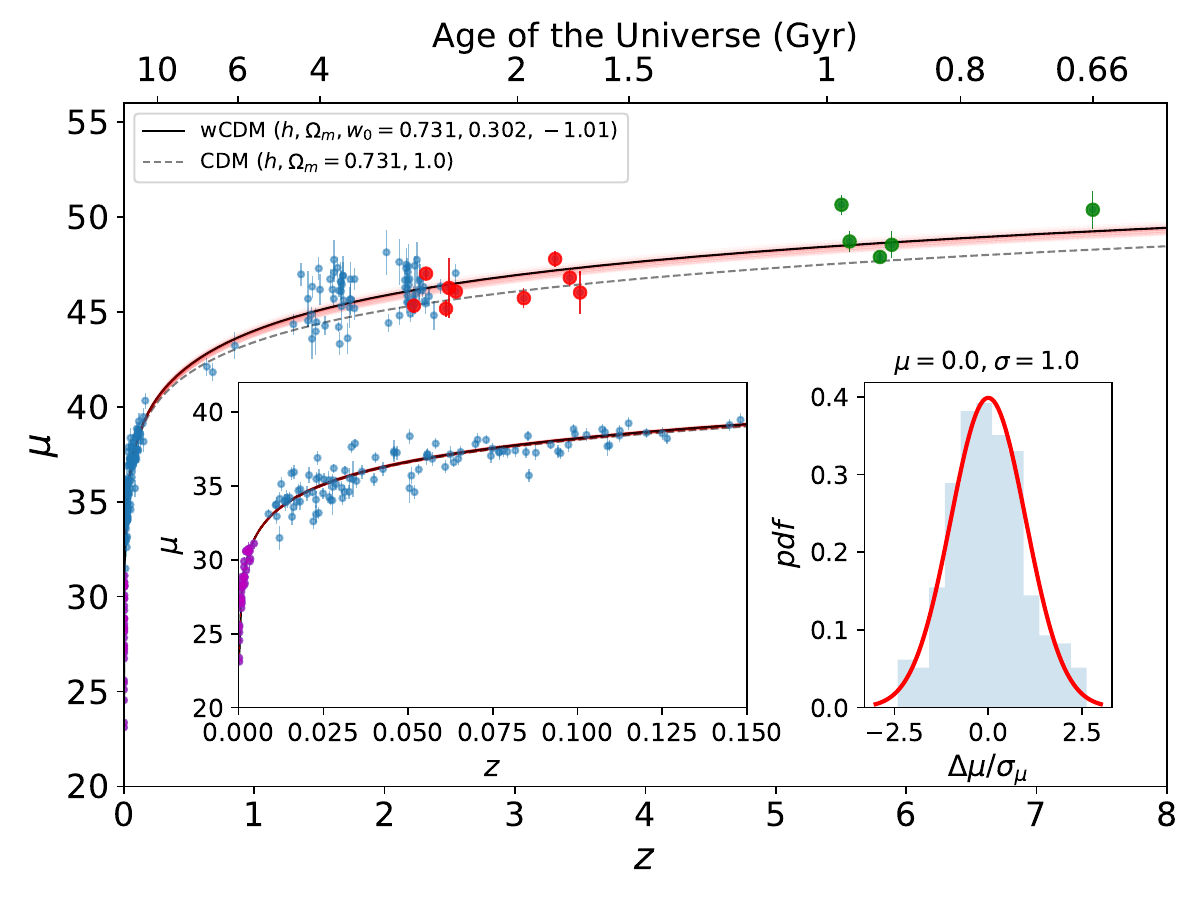}
\end{center}    
\caption{Hubble diagram for the HIIGs and anchor samples. $z$ is the redshift and $\mu$ is the distance modulus. In magenta we present the anchor sample of 36 objects which have been analysed in \citet{Fernandez2018}; in blue, the full sample of 181 HIIGs which have been analysed in \citet{2021MNRAS.505.1441G};  in red we present the 9 new HIIGs from \citet{2023A&A...676A..53L} and in green the 5 new HIIGs studied with JWST by \citet{2024A&A...684A..87D}. The inset at the left shows a close-up of the Hubble diagram for $z \leq 0.15$. The black line illustrate the cosmological model that best fit the data with the red shaded area representing the 1$\sigma$ uncertainties to the model, while the grey dashed line is a flat cosmological model without dark energy. The inset at the right presents the pulls probability density function (pdf) of the entire sample of GEHRs and HIIGs and the red line shows the best Gaussian fit to the pdf.}
\label{fig:hubdiag}
\end{figure*}

\begin{figure*}
\begin{center}
\includegraphics[width=1.6\columnwidth]{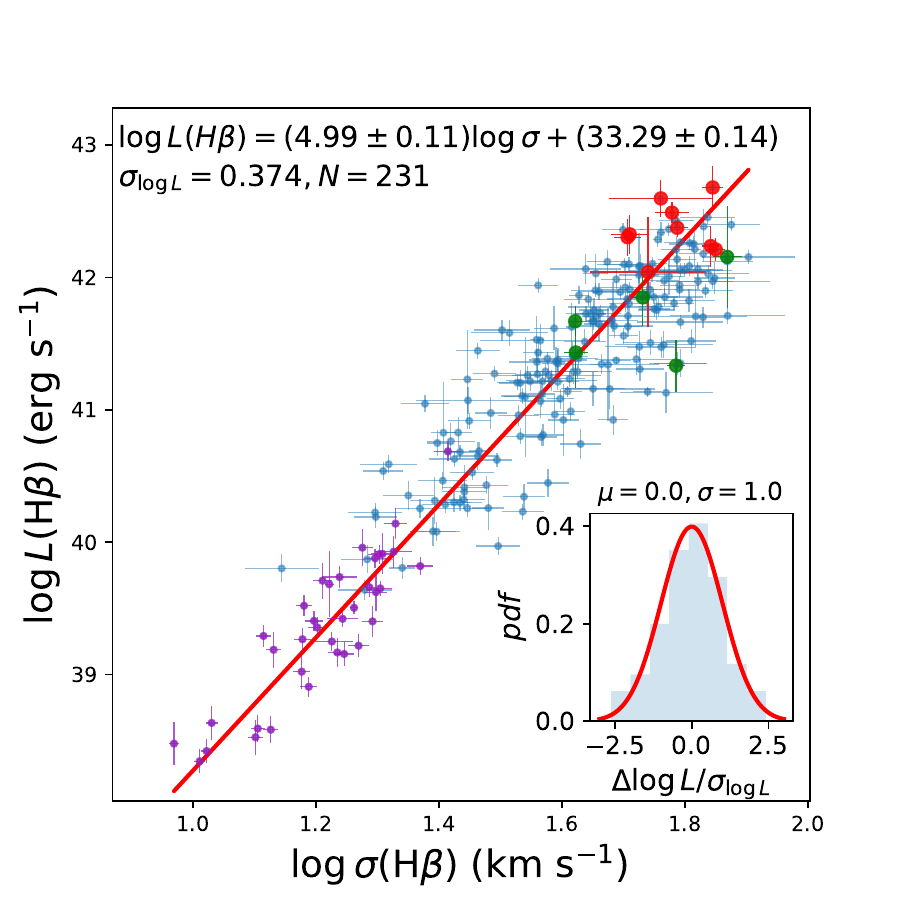}
\end{center}    
\caption{The \lsigG\ relation of the HIIGs and anchor samples. The data points follow the same colour code for the different samples as in the previous figure. 
The red line shows the best linear fit to the data, including the uncertainties in both axis. At the top of the figure we present the values of the slope and intercept of the best fit including their uncertainties. We also show the standard deviation of  $\log$ L around the best fit and the total number of objects in the sample. The inset shows the pulls distribution of the entire sample of GEHRs and HIIGs and the red line shows the best Gaussian fit to the distribution.}
\label{fig:LSig}
\end{figure*}

\begin{figure*}
\begin{center}
\includegraphics[width=1.6\columnwidth]{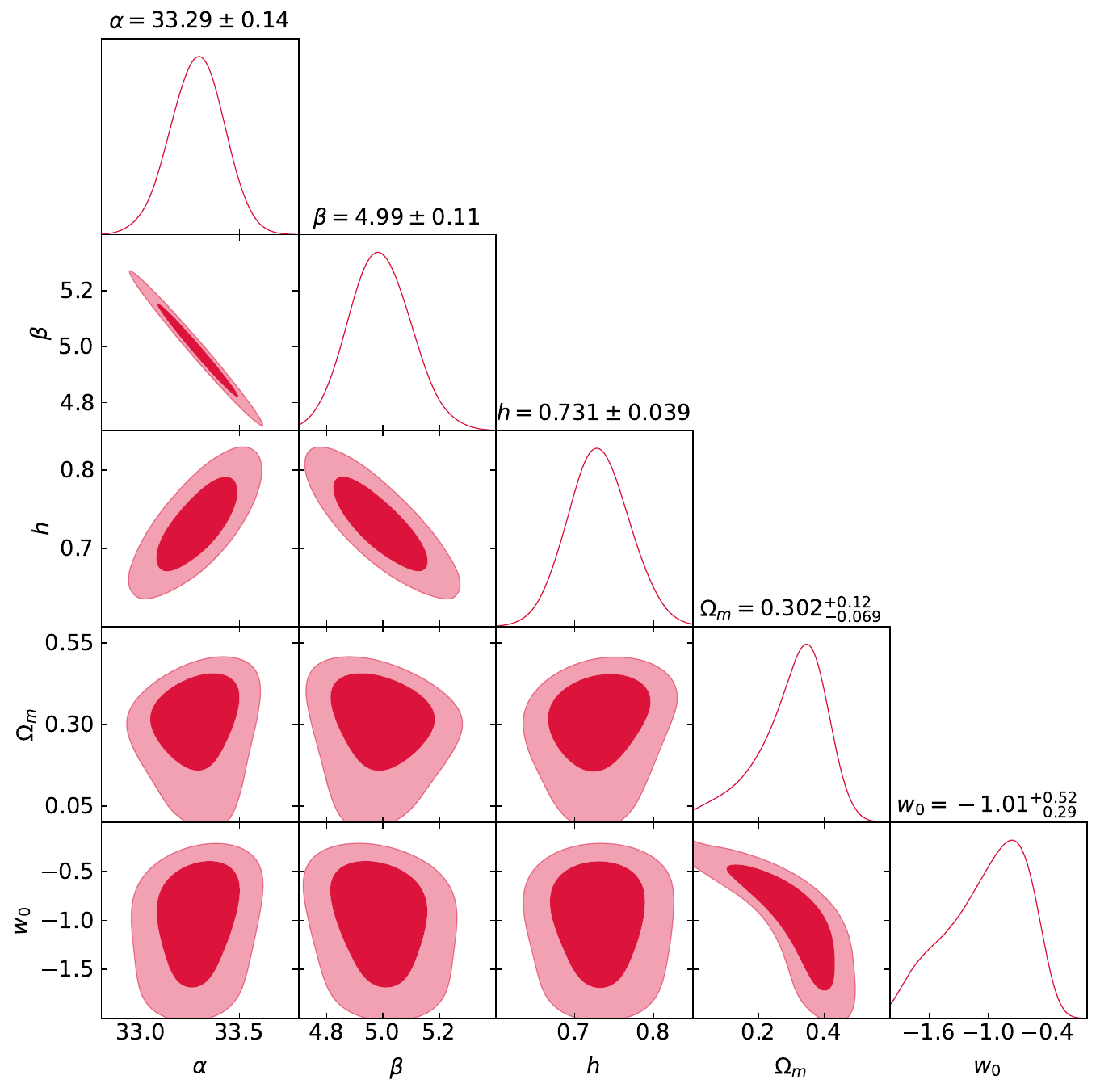}
\end{center}    
\caption{Likelihood contours corresponding to the 1$\sigma$ and 2$\sigma$ confidence levels in the $\{\alpha, \beta, h,\Omega_m\, w_0\}$ space for the joint HIIGs and anchor samples.}
\label{fig:abhOmW0}
\end{figure*}

\section{Results}
Using our compiled dataset of 231 objects from our HIIGs and anchor samples, we have derived constraints applicable to various cosmological models. 
The derived constraints on various cosmological and nuisance parameters are detailed in Table \ref{tab:const}. This table presents the marginalised best-fit values alongside their corresponding $1\sigma$ uncertainties for each parameter. Note that parameters enclosed in parentheses were held constant during the analysis, in this cases we adopted their values as in \citet{2021MNRAS.505.1441G}. Also, the table encompasses combined analyses of both HIIGs and our anchor samples, as well as scenarios where only the HIIGs sample is employed. 

Our main focus consists  on constraining a generalised parameter space, denoted as $\theta = \{\alpha, \beta, h, \Omega_m, w_0, w_a \}$. Here, $\theta_n = \{\alpha, \beta \}$ represents nuisance parameters, specifically characterising the \(\text{L}-\sigma\) relation for GEHRs and HIIGs, where $\alpha$ is the intercept and $\beta$ the slope of this relation. The remainder of the parameter space pertains to distinct cosmological models. For the flat $\Lambda$CDM model, the parameters are defined as $\theta_c = \{h, \Omega_m, -1, 0 \}$, indicating that we constrain the reduced Hubble constant $h = H_0/(100\ \mathrm{km\ s^{-1}\ Mpc^{-1}})$ and the total matter density parameter, $\Omega_m$, while maintaining constant the first two DE EoS parameters at $w_0 = -1$ and $w_a = 0$. This specific setting aligns with a cosmological constant 
($\Lambda$). Extending the constraints to include the value of $w_0$ allows for an evolving DE EoS, characteristic of models akin to quintessence \citep{PhysRevD.37.3406, Wetterich1988}. Lastly, incorporating a constraint on $w_a$ aligns with the Chevallier-Polarski-Linder (CPL) model, as delineated in the seminal works \citep{Chevallier2001, Linder2003, Peebles2003}.

In the first two rows of Table \ref{tab:const} we use our \( h \) free approach, so $\alpha$ and $h$ are not present in the analysis and we fix the value of $\beta$ as discussed above, so we do not employ the anchor sample data. In the following two rows of the table, we include constraints on $h$ but fix the values of the nuisance parameters, $\alpha$ and $\beta$, so again we do not employ the anchor sample. In the last four rows of the table, we show the results including the anchor sample, so that we are constraining simultaneously nuisance and cosmological parameters. The fact that in all cases the values of the nuisance parameters, $\alpha$ and $\beta$, are consistent at the level $1\sigma$ shows the stability of the analysis and the constraints on the \(\text{L}-\sigma\) relation.

In Figure \ref{fig:hubdiag} we present the Hubble diagram for the HIIGs and anchor samples. In magenta we plot the anchor sample of 36 objects which have been analysed in \cite{Fernandez2018}, in blue we present the full sample of 181 HIIGs  from \cite{2021MNRAS.505.1441G}, while in red we show the 9 new HIIGs from \cite{2023A&A...676A..53L} and in green 5 HIIGs newly observed with JWST from \cite{2024A&A...684A..87D}. The black line is the cosmological model that best fits the data with the red shaded area representing the 1$\sigma $ uncertainties to the model, while the grey dashed line is a flat cosmological model without dark energy. The inset at the left shows a close-up of the Hubble diagram for $z \leq 0.15$. The inset at the right presents the normalised residuals (or `pulls') distribution of the entire sample of GEHRs and HIIGs and the red line shows the best Gaussian fit to the distribution.

In Figure \ref{fig:LSig} we showcase the \(\text{L}-\sigma\) relation for the HIIGs and anchor samples. The data encompass four distinct groups, as explained
above.
The red line in the diagram represents the best linear fit to the data, accounting for uncertainties in both luminosity and velocity dispersion axes. Atop the figure we present the slope and intercept values of this best-fit line, along with their respective uncertainties. Additionally, we quantify the standard deviation of the logarithm of the luminosity (\(\log L\)) around the best fit, providing a measure of the scatter in the data. The total number of objects in the combined sample is also noted, offering a sense of the statistical robustness of the analysis. The inset in the figure displays the normalised residuals (or `pulls') distribution for the entire dataset. The best Gaussian fit to these residuals is represented by the red line showing that the pulls distribution follows closely the fit. This comprehensive analysis of the \(\text{L}-\sigma\) relation across a diverse set of GEHRs and HIIGs, including the latest JWST data, provides valuable insights into the underlying physics of these galaxies and contributes significantly to our understanding of galactic dynamics and star formation processes.

Figure \ref{fig:abhOmW0}  depicts  the 1$\sigma$ and 2$\sigma$ likelihood contours derived from a comprehensive global fit applied to our HIIGs and anchor samples. This fit encompasses all free parameters, both nuisance and cosmological, within the framework of a model featuring an evolving DE EoS parameter. The resultant parameter space constraints, as elucidated in this figure and also from Table \ref{tab:const}, demonstrate a high degree of consistency with other recent determinations in the field \citep{Scolnic2018, 2022ApJ...938..110B}. A comparative analysis of the Figure of Merit ($FoM$) with that reported in \citet{2021MNRAS.505.1441G} reveals a notable improvement of approximately 6.7\% in our results.

\section{Discussion and conclusions}
The observation that the \(\text{L}-\sigma\) relation remains valid to high-redshifts (\( z > 3 \)) HIIGs, extending to the epoch of reionization, unveils a remarkable uniformity in H II galaxy properties over vast cosmic timescales. This continuity is not just a testament to the robustness of the \( L-\sigma \) relation as a cosmological tool but also illuminates the fundamental processes governing formation and evolution of galaxies in the early Universe.

This result has also profound implications for our understanding of star formation processes in the early Universe. It suggests that the photo-kinematical properties  of massive young clusters, which ionise GEHRs and HIIGs, have remained unchanged for most of the age of the Universe. This conclusion  challenges models and assumptions about the non-universality of star formation mechanisms \citep{2010ARA&A..48..339B, 2022MNRAS.517.2471Z}.

The constraints on cosmological parameters deduced from our dataset, as delineated in Table \ref{tab:const} and Figure \ref{fig:abhOmW0}, specifically our constraints on the space $\lbrace h, \Omega_m, w_0\rbrace = \lbrace 0.731\pm0.039, 0.302^{+0.12}_{-0.069}, -1.01^{+0.52}_{-0.29}\rbrace $ (stat) from GHIIR and HIIG alone, are closely aligned with the latest results from the Pantheon+ analysis of 1550 SNIa $\lbrace h, \Omega_m, w_0\rbrace = \lbrace 0.735\pm0.011, 0.334\pm0.018, -0.90\pm0.14\rbrace $ \citep{2022ApJ...938..110B}. This concordance underscores the robustness and relevance of our findings in the broader context of contemporary cosmological research.

In their study of the \( L-\sigma \) relation for HIIGs as cosmological distance indicators, \citet{2024PhRvD.109l3527C} suggest that the relation exhibits a significantly flatter slope for high-$z$ HIIGs compared to their low-$z$ counterparts. This hypothesis carries substantial implications for employing HIIGs in determining cosmological distances. However, the flattening of the $L-\sigma$ slope at high-$z$ reported by \cite{2024PhRvD.109l3527C} is a direct consequence of the very restricted luminosity range accessible due to observational limits, which at high redshift predominantly capture the more luminous galaxies. This limited dynamical range in luminosity leads to a small value of the fitted slope, a concept that is supported by analyses indicating that the $L-\sigma$ slopes for high-$z$ and appropriately matched low-$z$ samples are statistically indistinguishable when comparing similar luminosity ranges. Additionally, the $L-\sigma$ relation may be affected by systematic variations in age, metallicity, and extinction corrections, necessitating meticulous control over these parameters in cosmological studies. Future research aimed at broadening the $L-\sigma$ application to a more diverse set of galaxy samples, including lensed GHIIRs and HIIGs (see \citet{2016A&A...592L...7T} for an example), could further elucidate the relation's validity across varied cosmic conditions, thereby confirming its universality and enhancing its utility in cosmology.

In our endeavour to refine the independent determination of cosmological parameters using HIIGs, the incorporation of additional data from the JWST up to and above $z \sim 9$ promises to be invaluable. The unparalleled sensitivity and resolution of JWST, capable of probing the early Universe, offer an unprecedented opportunity to observe HIIGs at higher redshifts. This extended observational reach is pivotal, as it allows for the exploration of the Universe's expansion dynamics under different cosmological conditions, thereby providing a more comprehensive understanding of the evolution of $\Omega_m$, the matter density parameter, and $w_0$, the dark energy equation of state parameter.

The observation of distant HIIGs with JWST not only enhances the statistical power of our analysis, but by enormously extending the accesible redshift range allows to test the consistency of the $\Lambda$CDM model  and the possible evolution of dark energy over a broader span. Moreover, this approach addresses potential biases inherent in current datasets predominantly stemming from their limited redshift range. The use of a single and independent distance indicator over an extremely wide range of cosmic history, with an increased data set is crucial for reducing statistical uncertainties and refining the constraints on $\lbrace h, \Omega_m, w_0\rbrace$.

Further observations from JWST will enable a more detailed examination of the intrinsic properties of HIIGs. This deeper insight is essential for the  calibration of the \(\text{L}-\sigma\) relation and possible mitigation of systematic uncertainties on the derived parameters. By enhancing our understanding of the physical processes governing HIIGs, we can better interpret their \(\text{L}-\sigma\) relation, a fundamental factor in the measurement of cosmological parameters.

Our analysis introduces novel insights into the evolution of the Universe. By leveraging this new data, we significantly enhance the existing narrative of cosmic history. Our results add to the understanding of the early Universe conditions and their role in the formation and evolution of galaxies setting the stage for expanding the exploration  of the photo-kinematical properties of  massive regions of star formation.

\section*{Data availability}
The datasets supporting the conclusions of this article, including the data used for generating the figures, are available in the references given at the Data Sets section and from the corresponding author upon reasonable request.

\section*{Code availability}
AstroPy \cite{astropy:2022}, Multinest \cite{Feroz2009, 2019OJAp....2E..10F} and PyMultinest \cite{2014A&A...564A.125B}, are all publicly available, while the code used for the data analysis and figure generation for this article is publicly available via GitHub at \url{https://github.com/blackdragonae/hiigs}

\section*{Acknowledgements}
RCh and DF-A acknowledge support from the CONAHCYT research grant CF2022-320152. RA acknowledges the support of project PID2023-147386NB-I00 and the Severo Ochoa grant CEX2021-001131-S funded by MCIN/AEI/10.13039/50110001103.
MLl acknowledges support from the PRIN 2022 MUR project 2022CB3PJ3 - First Light And Galaxy aSsembly (FLAGS) funded by the European Union – Next Generation EU.

\bibliography{bib/bib2022}
\label{lastpage}

\end{document}